# The Unique Citing Documents Journal Impact Factor (Uniq-JIF) as a Supplement for the standard Journal Impact Factor

Zhesi Shen[1,*], Li Li[1], Yu Liao[1]

National Science Library, Chinese Academy of Sciences, 100190, China

*shenzhs@mail.las.ac.cn

**A new type of impact factor**

The Unique citing documents Journal Impact Factor (Uniq-JIF) is defined as follows:

$$\text{Uniq-JIF} = \frac{\text{Number of unique citing documents}}{\text{Number of citable items}} \qquad (1)$$

We note that formula (1) is given in a generic form. In concrete applications, one must state the publication for which the Uniq-JIF is calculated (a journal, an edited book, a conference proceedings), the period during which citing occurs, the period during which the citable items are published, and which items are considered to be citable (this could be all items). As shown in Fig. 1, if one document cites journal A multiple times, it will only be counted once in the Uniq-JIF calculation. The idea of only counting unique citing documents can be traced back to at least Rousseau and Rons (2008). The Uniq-JIF of a journal measures an average number of unique citing documents per citable item published by this journal.

The rationale behind the Uniq-JIF is to provide a supplementary view to the traditional Journal Impact Factor (JIF). More specifically, the Uniq-JIF aims to reduce, but not eliminate, the influence of abnormal citation practices, such as citation manipulations, coercive self-citation, and citation stacking, that can artificially inflate the JIF of some journals. By focusing on the number of unique citing documents rather than total citations, the Uniq-JIF offers a more nuanced and fair representation of a journal's influence within the scientific community.

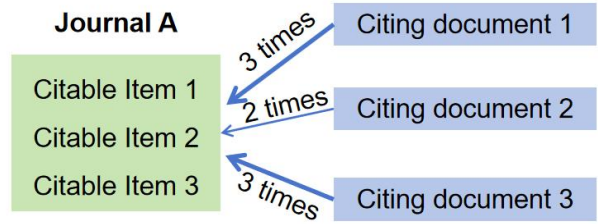

$$\text{Uniq-JIF} = \frac{1+1+1}{3} = 1$$

$$\text{JIF} = \frac{3+2+3}{3} = 2.67$$

Figure 1: Example of a Uniq-JIF calculation.

**What can be observed from the calculation of Uniq-JIFs?**

By analyzing the citation data provided in the recently released 2023 Journal Citation Reports, we calculate the Uniq-JIF and the ratio of Uniq-JIF to JIF for all the indexed SCI, SSCI, and ESCI journals. Here we used the same periods for the (Uniq-IF)(Y) as for the classical JIF, see (2), where Y refers to a fixed year. In this formula, UCIT(Y,{Y-1,Y-2}) refers to the number of unique documents citing in the year Y articles published in the years Y-1 or Y-2. Pub(X) denotes the number of articles (considered to be citable by Clarivate Analytics) published in the year X. Here X = Y-1 or Y-2.

$$(Uniq - JIF)(Y) = \frac{UCIT(Y, \{Y-1, Y-2\})}{PUB(Y-1) + PUB(Y-2)} \quad (2)$$

Note that if a document cites an article published in the year Y-1 and moreover another article published in the year Y-2 (in the same journal), it counts as a single unique citing document. This is a key distinction from the calculation of the standard Journal Impact Factor (JIF).

Figure 2 shows the distribution of the (Uniq-JIF)/JIF ratio. We see that for most journals, the drop of the Uniq-JIF compared to the JIF is less than 20%. However, we also observed 13 journals dropping more than 75%.

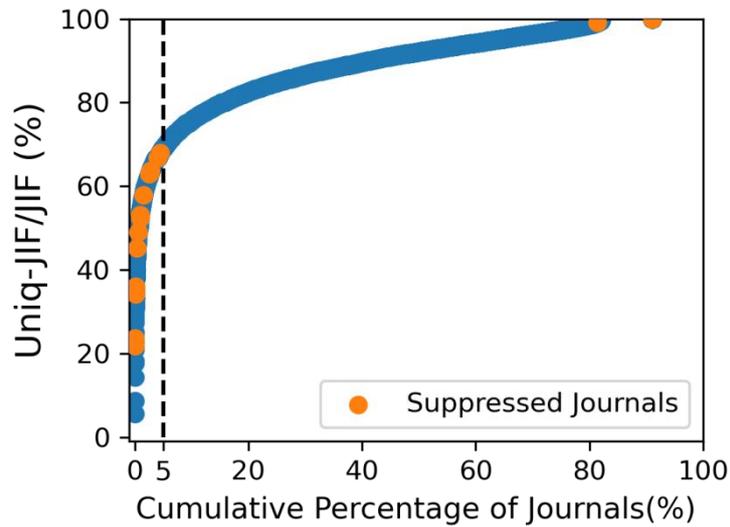

Figure 2. Cumulative distribution of the ratio (Uniq-JIF/JIF). Suppressed journals are represented by orange dots.

In addition, we calculate the Uniq-JIF for the 17 journals[1] that were suppressed due to citation-related issues (e.g., excessive self-citation, citation-stacking). As seen in Fig. 2 (look at the orange dots), 15 of these suppressed journals have a drop of more than 30%, placing them in the top 5% of journals with the largest fraction impact drop.

From this observation, we suggest that the Uniq-JIF may help in revealing potentially problematic journals.

**Conclusion**

This article introduces the Unique Citing Documents Impact Factor. Its analysis provides insights into the impact of citation-related issues on journal metrics and can help identify journals that may require further scrutiny.

**References**

JCR (2023). 2023 Journal Citation Reports

Rousseau, R., and Rons, N. (2008). Another *h*-type index for institutional evaluation. Current Science, 95(9), 1103.

---

[1] List of suppressed journals. Title Suppressions (clarivate.com)

**Conflict of Interest**

Zhesi Shen is a vice editor of the Journal of Data and Information Science.

**Statement**

During the preparation of this work the author(s) used LLM in order to improve readability and language. After using this tool/service, the author(s) reviewed and edited the content as needed and take(s) full responsibility for the content of the publication.